\documentclass[aps,prd,showpacs,preprintnumbers,nofootinbib,twocolumn]{revtex4}
\usepackage[dvips]{graphicx}
\usepackage{bm,latexsym,amsmath,amssymb,amsfonts}
\newcommand*{\B}{{{\rm b}}}
\newcommand*{\Bp}{{{\rm b},0}}
\newcommand*{\DMp}{{{\rm m},0}}
\newcommand*{\CH}{{\rm Ch}}

\newcommand*{\LS}{{\rm LS}}
\newcommand*{\TR}{{\rm tr}}
\newcommand*{\D}{{\rm d}}

\newcommand*{\ISW}{{\rm ISW}}
\newcommand*{\SW}{{\rm SW}}
\begin{document}
\title{A note on observational signatures
in superluminal unified dark matter models}

\author{Yuko~Urakawa}
\email[Email: ]{yuko"at"gravity.phys.waseda.ac.jp}
\author{Tsutomu~Kobayashi}
\email[Email: ]{tsutomu"at"gravity.phys.waseda.ac.jp}
\address{\,\\ \,\\
Department of Physics, Waseda University,
Okubo 3-4-1, Shinjuku, Tokyo 169-8555, Japan
}

\begin{abstract}
We explore the possibility that the dark matter and dark energy
are mimicked by a single fluid or by a single $k$-essence-like scalar field.
The so called Chaplygin gas unified dark matter models
can reproduce the observed matter power spectrum
by adding a baryon component.
It has been argued that the evolution of the baryon fluctuations
is particularly favoured for
the ``superluminal'' case where the sound speed of the Chaplygin gas
exceeds the speed of light at late times,
as well as for the models with the negligibly small sound speed.
In this note we compute the integrated Sachs-Wolfe signal
in the Chaplygin gas models, focusing on the superluminal case
which has not been investigated before because of the premature
understanding of causality.
It is shown that the superluminal model leads to large enhancement
of the integrated Sachs-Wolfe effect, which is inconsistent with the CMB measurements.
\end{abstract}

\preprint{WU-AP/302/09}
\maketitle

\section{Introduction}

Current cosmological observations indicate
that a large
fraction of the Universe is made of unknown ``dark'' components:
dark matter and dark energy. Their identities and physical properties are 
still masked in mystery.
Is dark energy a cosmological constant or an evolving scalar field?
Can modified gravity be an alternative to dark matter and/or dark energy?
Are dark matter and dark energy really distinct components?
In this note, this last possibility is explored.
Our goal is to reevaluate the viability of (a particular class of)
unified dark matter/dark energy models
from the point of view of cosmological observations.

Unification of dark matter and dark energy is an attractive idea
because it would be simpler than to consider two separate dark components.
Unified dark matter (UDM) models are often described by
a $k$-essence-like scalar field $\phi$~\cite{k-essence},
the Lagrangian of which is of the form
${\cal L}=p(X, \phi)$ with $X:=-(\partial\phi)^2/2$.
The scalar field is assumed to mimic both dark matter and dark
energy~\cite{Scherrer:2004au, Bertacca:2007ux, Bertacca:2007cv, Bertacca:2008uf}.
The basic requirements for a UDM model to be viable are:
(i) the background evolution is composed of the early-time matter-dominated phase
and late-time accelerated phase;
(ii) the evolution of cosmological perturbations
is consistent with observations.
Among various models we are interested in those which
are described by~\cite{Kamenshchik:2001cp, Bilic:2001cg, Bento:2002ps}
\begin{eqnarray}
 {\cal L} = - V_0 \left[ 1 - \left( X \over V_0 \right)^{(1 +\alpha)/2 \alpha } \right]^{ \alpha/(1 + \alpha)}
 \label{action:gDBI}
\end{eqnarray}
with $X<V_0$.
In the case of $\alpha=1$ this reduces to the Dirac-Born-Infeld Lagrangian
that appears in string theory~\cite{Alishahiha:2004eh, Silverstein:2003hf}, but
it is phenomenologically interesting to consider general $\alpha\,(\ge0)$.
The scalar field model~(\ref{action:gDBI}) has an equivalent description
in terms of a barotropic fluid with the equation of state
\begin{eqnarray}
p_\CH =-A \rho_\CH^{-\alpha}. \label{EOS:gCg}
\end{eqnarray}
The fluid version is referred to as the generalized Chaplygin gas (gCg)
model~\cite{Kamenshchik:2001cp, Bilic:2001cg, Bento:2002ps}.
Cosmological consequences of the gCg model have been extensively
investigated so far~\cite{Carturan:2002si, Gorini:2002kf, Sandvik:2002jz, Bean:2003ae, Colistete:2003xx,
Beca:2003an, Amendola:2003bz, Tonry:2003zg, Zhang:2005jj, Wu:2006pe,
Wu:2007bv, Davis:2007na, Fabris:2008mi, Fabris:2008hy, Batista:2009yu}.

The behavior of cosmological fluctuations
basically depends on the sound speed $c_s$ of a fluid,
and in the gCg case the sound speed crucially depends on the parameter $\alpha$. 
Thus, the allowed parameter space is
strongly constrained by the data on large scale structure (LSS)~\cite{York:2000gk, Abazajian:2008wr, Tegm} to be
$\alpha\ll 1$,
leaving the models which are essentially identical to the usual
cold dark matter model with a cosmological constant ($\Lambda$CDM), in particular
in the simplest modeling of the universe filled only with the gCg~\cite{Sandvik:2002jz}.
The situation seems to be drastically cured at least in the linear
regime by adding a baryon
component~\cite{Colistete:2003xx, Beca:2003an, Gorini:2007ta,Park:2009np},
as baryon fluctuations grow in a standard way to reproduce
the observed matter power spectrum consistently. \footnote{ While the Chaplygin gas
fails to cluster due to its increasing sound speed, the baryons can
cluster because of the absence of the direct interaction between the
Chaplygin gas and the baryons. This implies that an unusual bias is
required to relate the density fluctuations of these two components.}  
Since the linear growth factor of the
baryonic fluctuation is still suppressed to about $90\%$ compared
to the $\Lambda$CDM model, measurements of the cosmological
parameter $\sigma_8$ are expected to yield a strong constraint on the
gCg model~\cite{Sandvik:2002jz}. However, the limits from $\sigma_8$ are still
unclear because $\sigma_8$ is strongly affected by non-linear effects~\cite{Gorini:2007ta,Park:2009np}.


Most of the literature so far has focused on the models with $\alpha\le 1$
probably because the sound speed exceeds the light velocity for 
$\alpha > 1 $ ~\cite{Bonvin:2006vc, Bonvin:2007mw}.
However, as argued in~\cite{Babichev:2007dw}, the superluminal
propagation of a $k$-essence-type scalar field does not necessarily lead to the
causality violation.
Interestingly, the authors of Ref.~\cite{Gorini:2007ta} found recently that
the growth of perturbations is particularly
favoured for $\alpha\gtrsim 3$ as well as for very small $\alpha$ (see
also~\cite{Yang:2008hda}). In~\cite{Gorini:2007ta}, a particular
solution was proposed to assure the preservation of
causality.  Their point is that the 
necessary condition to preserve causality is
that the signal velocity does not exceed the light velocity in
vacuum~\cite{Brillouin}. Modifying the Lagrangian~(\ref{action:gDBI})
in an unobservable regime, they showed that the signal velocity
does not exceed the light velocity.


This note also focuses on the parameter range $\alpha> 1$.
We investigate the integrated Sachs-Wolfe (ISW) signal
in the case where the sound speed of the gCg can become superluminal,
in addition to the power spectrum of the baryon fluctuation studied in~\cite{Gorini:2007ta}.
It is known that a finite sound speed enhances the ISW effect~\cite{Bertacca:2007cv}, due to which
the gCg with $0.01<\alpha\le 1$ is excluded~\cite{Carturan:2002si, Bean:2003ae, Amendola:2003bz}.
We shall show that the superluminal models also yields large enhancement of the ISW effect,
which clearly inconsistent with observations. In contrast to weak
lensing, the linear calculation of the ISW effect is reliable, and hence
this is probably most convenient way of constraining the superluminal
gCg model.

\section{Basic equations} \label{model}


\subsection{The background evolution}

We consider the universe filled with the gCg and baryonic components.
The two components are assumed to interact only gravitationally.
As was pointed out in~\cite{Beca:2003an, Amendola:2003bz, Gorini:2007ta},
it is important to
include the baryons to reproduce the observed matter power spectrum.
Since the
modification made to cure the causality problem does
not affect the cosmic evolution until now, we use 
the equation of state~(\ref{EOS:gCg}).
It follows from Eq.~(\ref{EOS:gCg}) and the energy conservation equation,
$\D \rho_\CH/\D t+ 3H(\rho_\CH+p_\CH)=0$, that
\begin{eqnarray}
\rho_\CH= \rho_{{\rm Ch},0} \left[\bar A+\frac{1-\bar A}{a^{3(1+\alpha)}}\right]^{1/(1+\alpha)}, 
\label{rhoCh}
\end{eqnarray}
where $t$ is the proper time, $a(t)$ is the scale factor of the universe, and
$H:=\D\ln a/\D t$ is the Hubble parameter.
The present value of the scale factor $a_0$ is set to unity.
$\rho_{{\rm Ch},0}$ is essentially an integration constant and
represents the present energy density of the gCg,
and $\bar{A} := A / \rho_{{\rm Ch},0}^{1 + \alpha}$.
The ratio between the pressure and energy density, $w_\CH$, is given by
\begin{eqnarray}
w_\CH:=\frac{p_\CH}{\rho_\CH} =-\left[
1+ \frac{1 - \bar{A}}{\bar{A}} \frac{1}{a^{3(1+\alpha)}}
\right]^{-1} .\label{w:gCg}
\end{eqnarray}
Equation~(\ref{w:gCg}) clearly shows that the gCg indeed mimicks both
the dark matter and dark energy: we have the matter dominant phase,
$w_\CH \simeq 0$, at early times and the dark energy phase,
$w_\CH \simeq -1$, at late times.
The transition from the matter dominant phase to the accelerated phase
proceeds more rapidly for larger $\alpha$.
(The transition time is determined through the model parameters $\alpha$ and 
$A$, as explained shortly.)
Note that in the limit $\alpha\to 0$ the gCg represents
usual dark matter plus a cosmological constant.

Assuming the spatially flat universe, the Friedmann equation can be recast in
\begin{eqnarray}
\frac{H^2}{H_0^2}=\frac{\Omega_\Bp}{a^3}
+(1-\Omega_\Bp)\left[\bar A+\frac{1-\bar A}{a^{3(1+\alpha)}}\right]^{1/(1+\alpha)},
\end{eqnarray}
where $H_0 = 100 \,h^{-1}\, {\rm km}\, {\rm s}^{-1}\, {\rm Mpc}^{-1}$
is the present Hubble parameter and $\Omega_\Bp$ is the
present baryon density fraction. 
Let us define the transition time as the time at which the deceleration parameter becomes zero.
The redshift at the transition time is then given by
\begin{eqnarray}
 { \bar{A} [(1 + z_\TR)^{-3 (1 + \alpha)} + 1/2 ] - 1/2 \over
 \{ \bar{A}[(1 + z_\TR)^{- 3 ( 1 + \alpha)} - 1] + 1 \} ^{\alpha/( 1+ \alpha )}}
 =   {\Omega_\Bp \over 2( 1 - \Omega_\Bp) }.
\end{eqnarray}
Instead, to avoid the complication
we may use the relation
\begin{eqnarray}
\bar A = \frac{(1+z_\TR)^{3(1+\alpha)}}{2+(1+z_\TR)^{3(1+\alpha)}}\label{Rel:barA}
\end{eqnarray}
to define the transition redshift. Since $\Omega_\Bp$ is very small,
the two definitions make practically no difference.
Let us then define the present value of the total (non-relativistic) matter density, $\Omega_\DMp$.
Noting that
\begin{eqnarray}
\frac{H^2}{H_0^2} \to\frac{\Omega_\Bp+(1-\Omega_\Bp)(1-\bar A)^{1/(1+\alpha)}}{a^3}
\end{eqnarray}
for $a\to 0$, we may read off from this that
\begin{eqnarray}
\Omega_\DMp:=\Omega_\Bp+(1-\Omega_\Bp)(1-\bar A)^{1/(1+\alpha)}.
\end{eqnarray}

\begin{figure}
\begin{center}
\includegraphics[width=7.8cm]{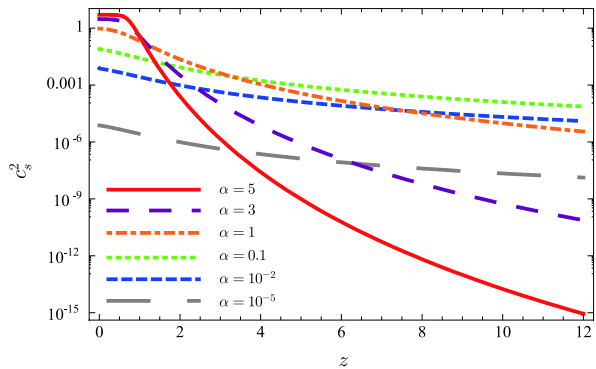}
\caption{The sound speed as a function of the redshift z
for $z_{{\rm tr}}=0.8$ and different $\alpha$.
The models with $\alpha \gtrsim 1$ show the
nontrivial behavior:
$c_s^2$ is smaller for larger $\alpha$ in the past,
which rapidly grows at $z\sim 1$ to be $c_s^2\sim \alpha$ at $z=0$.}  
\label{fg:cs2}
\end{center}
\end{figure}

Finally, we introduce the key quantity in the present note: the sound speed of the gCg,
\begin{eqnarray}
 c^2_s := \frac{\D \rho_\CH}{\D \rho_\CH} = - \alpha w_\CH .
 \label{cs:gCg}
\end{eqnarray}
The sound speed plays a crucial role in the following perturbation analysis.
The behavior of $c_s^2$ is plotted in Fig.~\ref{fg:cs2}.
The sound speed remains small before the transition time.
For $\alpha\gtrsim 1$, $c_s^2$ is particularly small in this regime.
It starts to grow
around the transition, and
$c_s^2$ can exceed the speed of light at late times for $\alpha>1$.
We thus call the gCg with $\alpha>1$ ``superluminal,''
though $c_s^2$ can be larger than 1 only at late times
and the model with $\alpha\simeq 1$ is in fact ``subluminal'' because $|w_\CH |<1$.

\subsection{Cosmological perturbations in the gCg model}

In the linear analysis, we adopt the longitudinal gauge
in which the metric is written as
\begin{eqnarray}
\D s^2 = -(1+2\Phi)\D t^2+a^2(1-2\Phi)\delta_{ij}\D x^i\D x^j .
\end{eqnarray}
Here, we neglect the anisotropic stress so that
the gravitational potential coincides with the
curvature perturbation in this gauge.
We do not consider the entropy perturbation,
and write the pressure perturbation of the gCg as
\begin{eqnarray}
\delta p_\CH =  c_s^2 \delta \rho_\CH, 
\end{eqnarray}
where $\delta \rho_\CH$ is the density perturbation of the gCg.

Combining the $(t ,t)$ and $(i, j)$ components of the perturbed Einstein equations
to eliminate $\delta\rho_\CH$,
we obtain (in the Fourier space)
\begin{eqnarray}
&&\Phi''+\left[4+3c^2_s +(\ln H)'\right]\Phi'+\left[3(1+c^2_s)+2(\ln H)'\right]\Phi
\nonumber\\&&\qquad
 +  \frac{c^2_sk^2}{a^2H^2}\Phi= - \frac{3}{2} c^2_s
 \frac{\rho_\B}{\rho_\B + \rho_\CH }\delta_\B,
 \label{EOM:Phi}
\end{eqnarray}
where $\delta_\B:=\delta\rho_\B/\rho_\B$ is the baryon density perturbation and
the prime denotes the differentiation with respect to $\ln a$. For brevity, we
abbreviate the suffix $k$ if not necessary.
The energy-momentum conservation of the baryonic component implies
\begin{eqnarray}
&&\left(\delta_\B-3\Phi\right)'=\frac{k^2}{H}v_\B,
\\
&&\left(a^2v_\B\right)'+\frac{\Phi}{H}=0,
\end{eqnarray}
where $v_\B$ is the velocity perturbation of the baryon fluid. 
These two equations are combined to give
\begin{eqnarray}
&&\delta_\B''+\left[2+(\ln H)'\right]\delta_\B' + \frac{9}{2} c_s^2  \frac{\rho_\B}{\rho_\B+\rho_\CH}\delta_\B
\nonumber\\&&\quad
=-3(2 + 3 c_s^2 )\Phi'
-3\left[3(1 + c_s^2 )+2(\ln H)'\right]\Phi
\nonumber\\&&\qquad
-\frac{k^2}{a^2H^2}(1 + 3 c_s^2 )\Phi, \label{EOM:deltab}
\end{eqnarray}
where we used Eq.~(\ref{EOM:Phi}) to eliminate $\Phi''$.
Equations~(\ref{EOM:Phi}) and~(\ref{EOM:deltab}) are sufficient for
predicting the ISW effect and matter power spectrum
in the gCg model.




\section{Observational signatures}  \label{result}

\subsection{Evolution of cosmological perturbations}

\begin{figure}
\begin{center}
\includegraphics[width=7.8cm]{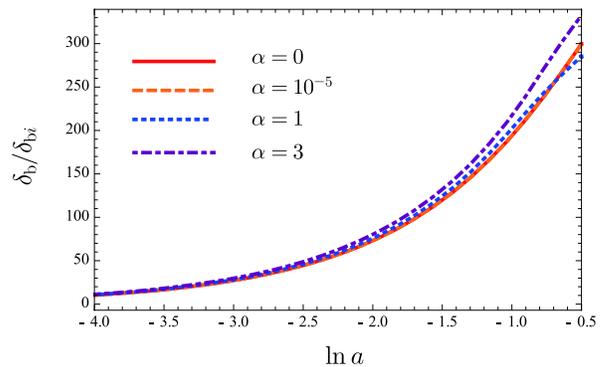}
\caption{The evolution of the baryon density perturbation
for $k=0.01\, h \,{\rm  Mpc}^{-1}$. The parameter of the background is fixed by
choosing $z_{{\rm tr}}=0.8$ and $\Omega_\Bp = 0.04$.
The evolution is not sensitive to $\alpha$.}  
\label{fg:deltab}
\end{center}
\end{figure} 

Let us begin with over-viewing the evolution of
the metric perturbation $\Phi$ and the baryon density perturbation $\delta_\B$
governed by Eqs.~(\ref{EOM:Phi}) and~(\ref{EOM:deltab}).
At early times we have $w_\CH\simeq 0$ and hence $c_s^2\simeq 0$.
Thus, for $z \gg z_\TR$, we can find the same solution as in the CDM
dominated universe. 
Noting that $H^2\propto a^{-3}$ in this limit, Eqs.~(\ref{EOM:Phi}) and~(\ref{EOM:deltab})
are solved to give
\begin{eqnarray}
\Phi \simeq C_k,
\quad
\delta_\B\simeq C_k\left[
-2-\frac{2}{3}\left(\frac{k}{a_i H_i}\right)^2 \frac{a}{a_i}
\right],
\end{eqnarray}
where the decaying mode has been ignored.
Here, $C_k$ is a constant, and $a_i$ and $H_i$ are the quantities evaluated at some initial time.
In practice, we may impose the initial condition at the last scattering surface:
$a_i=(1+z_{{\rm LS}})^{-1}\approx 10^{-3}$.
The primordial power spectrum is encoded in the $k$-dependent constant $C_k$.
Having specified the initial condition,
we separate the evolution of perturbations for $a>a_i$ from
the initial amplitude $C_k$ and
write the solution to Eqs.~(\ref{EOM:Phi}) and~(\ref{EOM:deltab})
in terms of the transfer functions as
\begin{eqnarray}
\Phi_k(a) = C_k {\cal T}_\Phi(k, a),
\quad
\delta_{\B k}(a) = C_k {\cal T}_\B(k, a).
\end{eqnarray}

The most important ingredient that determines the behavior of fluctuations
is the magnitude of the sound velocity $c_s^2$, or, equivalently,
the Jeans length of the gCg fluid: $\lambda_J^2:=c_s^2/(aH)^2$.
The gCg fluctuations with $k \gtrsim \lambda_J^{-1}$ are supported by the pressure
and oscillate, rather than grow.
This fact puts a very stringent constraint
on the unified dark matter model composed of a single gCg fluid,
as such oscillations hinder the gCg component to reproduce
the observed matter power spectrum.
One can, however, overcome this early drawback by adding
a baryon component, because
the baryon fluctuations are not affected much by the late-time oscillation
of the other component.
In other words,
the gravitational interaction is not strong enough to eliminate
the baryon fluctuations built up before the oscillatory regime.
The typical evolution of the baryon density perturbation is shown in Fig.~\ref{fg:deltab}.
Although the different choices of $\alpha$ lead to the different magnitude
of the gCg sound speed at late times,
$\delta_\B$ shows a very similar behavior, irrespective of $\alpha$.
One should note here that
even in the superluminal case with $\alpha>1$
the baryons can cluster because the sound speed remains small at early times.
Interestingly, the sound speed at early times
is much smaller for $\alpha>{\cal O}(1)$
than for $\alpha \lesssim  {\cal O}(1)$, and hence the former is much closer to
the standard $\Lambda$CDM model than the latter, at least at early times.

\begin{figure}[tb]
  \begin{center}
    \includegraphics[keepaspectratio=true,height=75mm]{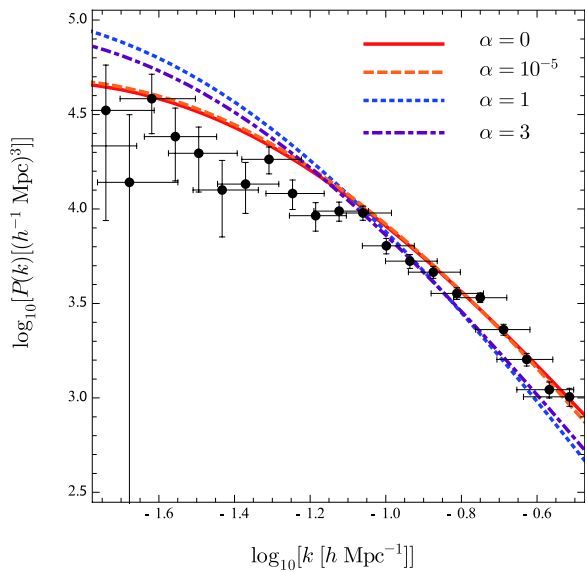}
  \end{center}
  \caption{The power spectra of the baryon component for different $\alpha$ and $z_\TR=0.8$~\cite{Gorini:2007ta},
  compared to the observed one~\cite{Tegm}.}%
  \label{fig:mtps.eps}
\end{figure}

In Ref.~\cite{Gorini:2007ta} Gorini {\em et al.} solved
the evolution of $\delta_\B$ and computed
the matter power spectrum, focusing in particular on the superluminal gCg model.
They argue that the growth of baryon inhomogeneities is observationally favoured
for $\alpha\gtrsim 3$ as well as for sufficiently small values of $\alpha$.
For the sake of completeness, we have repeated essentially the same calculation
as in~\cite{Gorini:2007ta}. The power spectra of the baryon component are shown
in Fig.~\ref{fig:mtps.eps}, which are in good agreement with the observed one.
To rely on the linear analysis, we discard the data with
$k > 0.3 h {\rm Mpc}^{-1}$.


\begin{figure}
\begin{center}
\includegraphics[width=7.8cm]{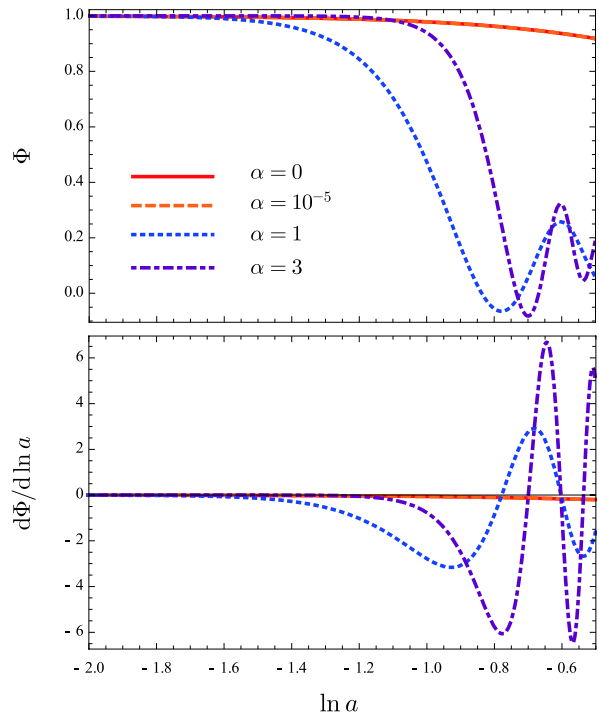}
\caption{The evolution of the metric perturbation $\Phi$
for $k=0.01\, h\, {\rm Mpc}^{-1}$. The parameter of the background is fixed by
choosing $z_{{\rm tr}}=0.8$ and $\Omega_\Bp = 0.04$.
In contrast to the baryon density perturbation,
the evolution of $\Phi$ crucially depends on $\alpha$.}  
\label{fg:phi}
\end{center}
\end{figure} 

Now we turn to the evolution of the gravitational potential $\Phi$.
In contrast to the baryon fluctuations,
$\Phi$ is crucially affected by the late-time increase in $c_s^2$.
As shown in Fig.~\ref{fg:phi},
the pressure-supported oscillation appears for
large values of $\alpha$,
leading to a strong time variation of $\Phi$.
This is simply because $c_s^2$ is larger for larger $\alpha$ at late times.
Let us focus on the cases with $\alpha=1$ and $\alpha=3$ and
compare them in more detail.
Since the sound speed remains $c_s^2 <(aH/k)^2$
for shorter time for $\alpha=1$ than for $\alpha=3$,
$\Phi$ begins to oscillate earlier in the former case.
In the accelerated phase, $z\lesssim z_\TR$,
the sound speed of the $\alpha=3$ model
grows to become larger than that of the $\alpha=1$ model,
which causes the larger time variation for $\alpha=3$ than for $\alpha=1$
in the oscillatory regime.

\subsection{ISW effect}

Having thus emphasized the characteristic behavior of $\Phi$,
it is important to study the signatures of observations
that probe more directly the gravitational potential itself.
An example is weak gravitational lensing, using which
the authors of~\cite{Sandvik:2002jz} claimed that the allowed parameter space
of the gCg model (with baryons) is very tiny. 
However, their argument is based on the linear analysis despite the fact that
nonlinear effects are important.
Another nice example is the ISW effect~\cite{Sachs:1967er, Rees:1968zz} .
A strong time variation of the gravitational potential at $z\lesssim z_\TR$
typically gives rise to large enhancement of the ISW signal.
This can put a stringent constraint on the model parameters,
and indeed the gCg models with $0.01<\alpha \le 1$
are excluded~\cite{Carturan:2002si, Bean:2003ae, Amendola:2003bz} (see
also~\cite{Bertacca:2007cv}).
In this note we
choose to pursue this direction, paying a particular attention to
the superluminal gCg with $\alpha>1$.

Noting that the gravitational potential coincides with the curvature
perturbation in the absence of anisotropic pressure,  the angular power
spectrum of the ISW effect is, for the case with a spatially flat
background, given by (see \cite{Bertacca:2007cv, Hu:1995em, book:Weinberg08, book:LiddleLyth})
\begin{eqnarray}
C_l^{{\rm ISW}}=4\pi T_0^2 \int\frac{\D
 k}{k} \frac{k^3|C_k|^2}{2\pi^2} \left[I_l^\Phi(k)\right]^2,
\end{eqnarray}
where
\begin{eqnarray}
I_l^\Phi(k) := 2 \int^{\eta_0}_{\eta_\LS}
 \partial_\eta {\cal T}_\Phi[k, a(\eta)] \, j_l[k (\eta_0 - \eta  )] \D \eta, 
 \label{Def:IPhi}
\end{eqnarray}
$j_l(x)$ is the spherical Bessel function,
and $\eta$ is the conformal time.
For our purpose, we focus on the late time ISW effect. 
The initial amplitude $C_k$ is obtained from the primordial spectrum multiplied
by the BBKS transfer function~\cite{Bardeen:1985tr}:
\begin{eqnarray}
\frac{k^3|C_k|^2}{2\pi^2} = \frac{9}{25}\Delta_{{\cal R}}^2(k) T^2(k),
\end{eqnarray}
with
\begin{eqnarray}
&&T(k)=\frac{\ln(1+2.34q)}{2.34q}
\nonumber\\&& \times\!
\left[1+3.89q+(16.1q)^2+(5.46q)^3+(6.71q)^4 \right]^{-1/4}\!, \nonumber \\
\end{eqnarray}
where $q:=k/(\Omega_{X,0}h^2)$ and $\Omega_{X,0}$ incorporates
Sugiyama's shape correction~\cite{Sugiyama:1994ed}:
\begin{eqnarray}
 \Omega_{X,0} :=  \Omega_{\DMp} \exp
 \left( - \Omega_\Bp - \frac{\sqrt{2h} \, \Omega_\Bp}{\Omega_\DMp}  \right) .
\end{eqnarray}
We simply assume that the primordial power spectrum
is the scale invariant Harrison-Zel'dovich one: $\Delta_{{\cal R}}^2(k)=$ constant.

\begin{figure}
\begin{center}
\includegraphics[width=8.5cm]{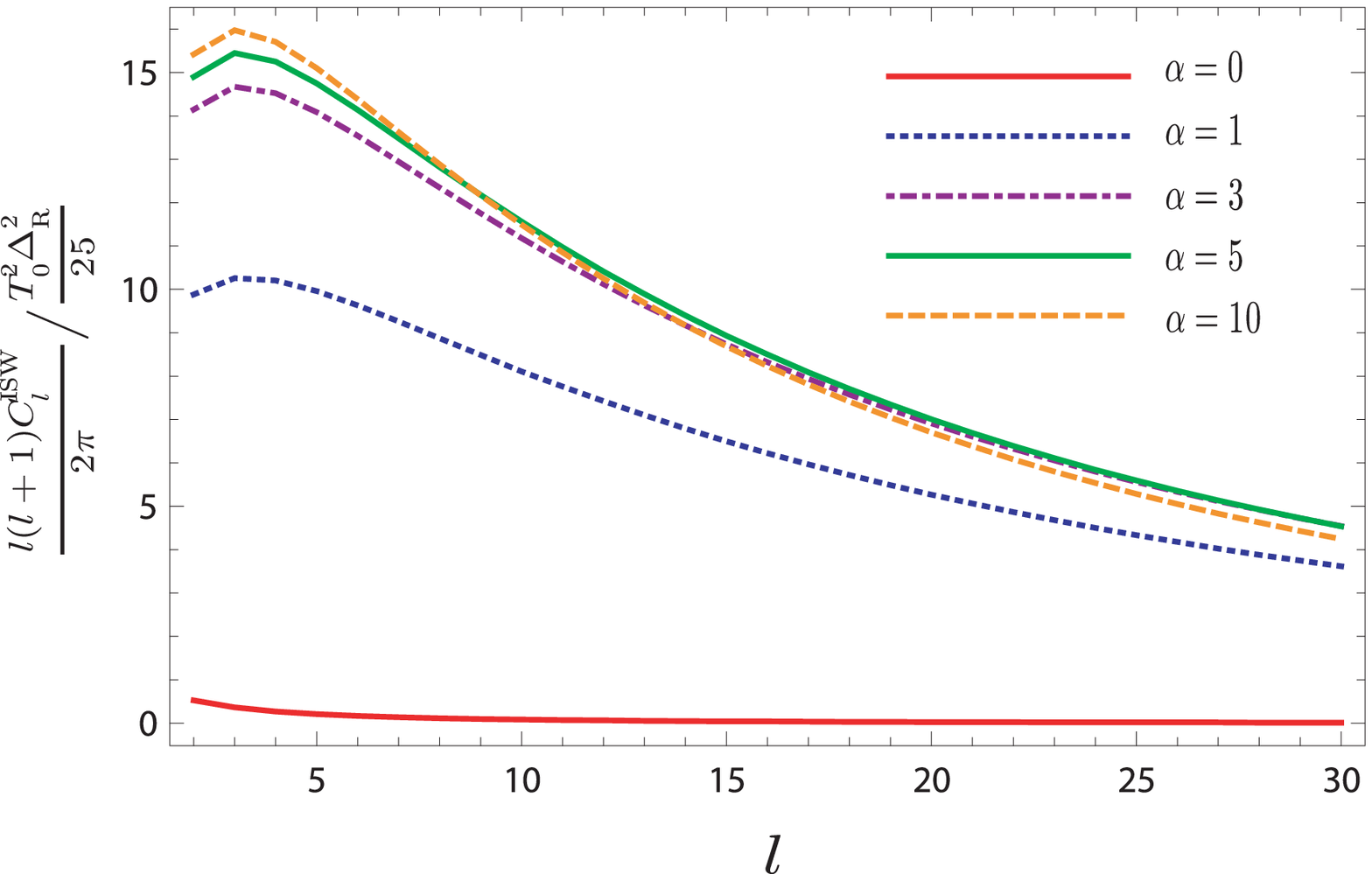}
\includegraphics[width=8.5cm]{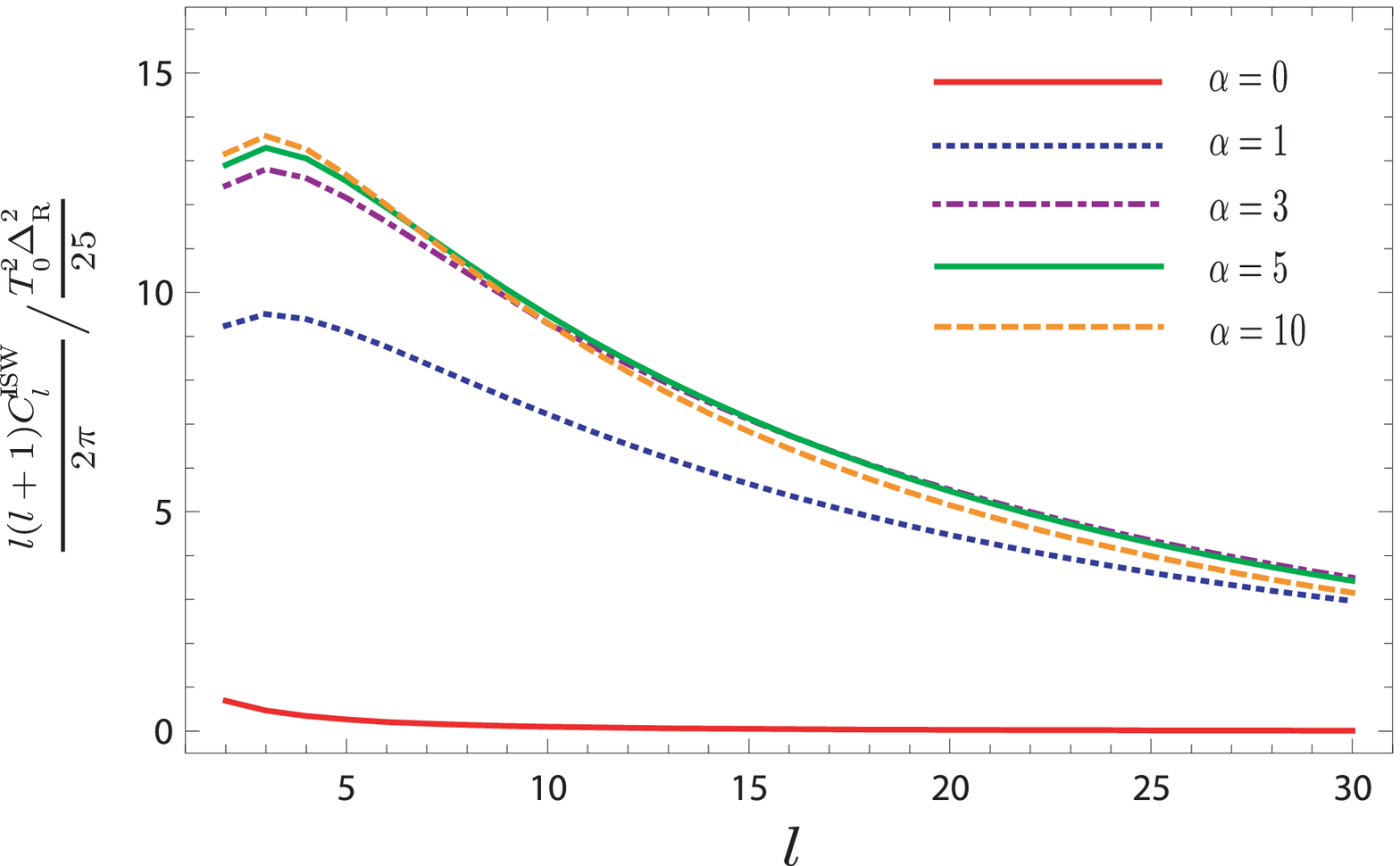}
\caption{The power spectra of the ISW effect normalized by 
$ T_0^2 \, \Delta^2_{{\cal R}} / 25$, which approximately gives the
angular spectrum for the SW effect. top: $z_\TR = 0.8$,
 bottom: $z_\TR=1.0$.}  
\label{fg:ISW/SW}
\end{center}
\end{figure} 

The Hubble Space Telescope Key Project \cite{Freedman:2000cf} result and
the WMAP 5yr data~\cite{Hinshaw:2008kr} tightly constrain the parameters
$h$ and $\Omega_\Bp$.
Since the amplitude of the ISW effect varies at most few percents within
the parameter region allowed by observations, we use
$h= 0.7$ and $\Omega_\Bp = 0.04$.
The angular spectra of the ISW effect for different $\alpha$ and $z_\TR$
are shown in Fig. \ref{fg:ISW/SW}. The amplitude is normalized by 
$ T_0^2 \, \Delta^2_{{\cal R}} / 25$, which approximately gives the amplitude of
angular spectrum for the Sachs-Wolfe effect $C_l^{\SW}$.
For the standard $\Lambda$CDM model, $\alpha=0$, 
$C_l^\ISW$ is roughly $10 \%$ of $C_l^{\SW}$. Therefore the
amplitude of the flat plateau is essentially determined by $C_l^{\SW}$.
Contrary, in the gCg model with $\alpha \gtrsim {\cal O}(1)$, $C_l^{\ISW}$ is more than ten times
larger than $C_l^{\SW}$, and hence the
amplitude of the flat plateau is essentially determined by $C_l^{\ISW}$.
Reflecting the oscillating behaviour of $\Phi$,
the enhancement of the ISW effect appears also in the superluminal model
with $\alpha > 1$. Since the baryon density perturbation $\delta_\B$ does not exhibit the distinctive
behaviour in the superluminal gCg model, this enhancement would help us to
discriminate this model.

\begin{figure}
\begin{center}
\includegraphics[width=8.5cm]{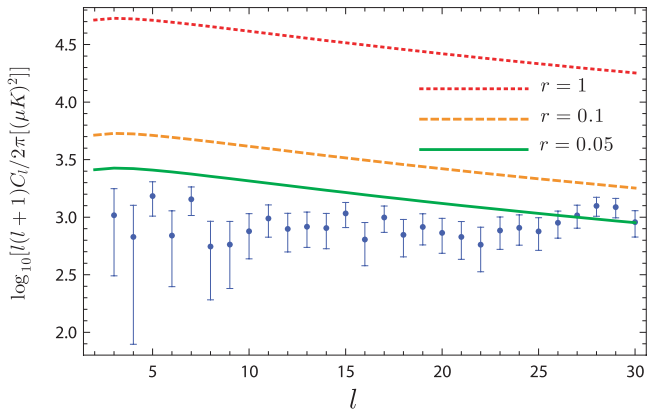}
\caption{The power spectra for $\alpha=3$, $z_\TR = 0.8$, and different
values of $\Delta^2_{{\cal R}}$, compared to the WMAP 5yr
 data~\cite{Hinshaw:2008kr}. $r$ is defined by
 $r := (\Delta_{{\cal R}}\,/\,\Delta^{\scriptsize{\rm{LSS}}}_{{\cal R}})^2$.}  
\label{fg:variousamp}
\end{center}
\end{figure}

Combining the constraints from the CMB anisotropies and the LSS, we
now discuss the viability of the superluminal gCg model. 
From the above argument,
one sees that, regarding the ISW effect,
a significantly smaller value of the primordial amplitude $\Delta^2_{{\cal R}}$
is favoured in the superluminal gCg model compared to the standard value in
the $\Lambda$CDM model.
However, in regards to the structure
formation, the preferred value of $\Delta^2_{{\cal R}}$
is larger by an order of magnitude in the
superluminal gCg model than in the $\Lambda$CDM model, since the
structure formation proceeds less efficiently in the gCg model. 
When we change the primordial amplitude
$\Delta^{\scriptsize{\rm{LSS}}}_{{\cal R}}\,^2$ 
to $r \Delta^{\scriptsize{\rm{LSS}}}_{{\cal R}}\,^2$, the angular spectrum $C_l$ and the matter power
spectrum $P(k)$ also change into $r C_l$ and $r P(k)$, respectively. 
Here, we define $\Delta^{\scriptsize{\rm{LSS}}}_{{\cal R}}\,^2$  as the prefered value to
reproduce the matter power spectrum, which depends on $\alpha$ and $z_\TR$. 
We need $r \lesssim 0.1$
in order for the temperature anisotropy to be compatible with observations,
but
then the predicted matter spectrum
significantly deviates from the SDSS data. 
As an example,
the total angular spectra $C_l$ for $r=1,\, 0.1$, and $0.05$
are shown in Fig.~\ref{fg:variousamp}. For $r= 0.1$
the amplitude of the matter power spectrum comes to be inconsistent with
observations, but even in this case the angular spectrum of the CMB still remains larger 
than the observed one. 

Finally, let us note the dependence on the parameters
$\alpha$ and $z_\TR$. We have computed the ISW signal for 
$\alpha \leq 10$ and found that these models are inconsistent with observations. 
The angular power spectrum for the ISW effect
and the matter power spectrum slightly depend on the transition time
$z_\TR$. Our result is robust to change of $z_\TR$ within the
observationally allowed region (which is not so small)~\cite{Melchiorri:2007in, Wu:2007bv}. Indeed, for $0.4 \leq z_\TR \leq 1.2$,
the superluminal gCg model with $\alpha \leq 10$ cannot be consistent 
both with the LSS
and with the CMB anisotropies at the same time.

\section{Conclusions}  \label{conclusion}

We have investigated the growth of inhomogeneities in
superluminal gCg cosmology, in which
a single fluid component mimics both dark matter and dark energy.
The superluminal gCg model is interesting observationally because
it can reproduce the observed matter power spectrum well and theoretically because
it does not in fact suffer from the causality problem.
We have shown that the late-time increase of the sound speed
affects the evolution of the metric potential $\Phi$ rather than that of
the baryon density perturbation $\delta_\B$. Since the time
dependent gravitational potential generates secondary temperature fluctuations
through the ISW effect, the observation of the CMB angular spectrum at
low multipoles helps to discriminate the superluminal gCg model. 
We have found that, due to large enhancement of the ISW effect, the superluminal
gCg model cannot explain the SDSS data and WMAP data consistently. Here,
we focused only on the flat plateau of CMB angular power spectrum,
combined with the matter power spectrum. As is the case in the
subluminal gCg model~\cite{Bean:2003ae, Amendola:2003bz}, normalizing
$C_l$ to COBE at $l=10$, it would be possible to
give stringent constraints from the acoustic peak of CMB, too.


In this note we have only considered a particular class of UDM models.
Then, a question is:
within the context of UDM models, how generic is large enhancement of the ISW effect?
It is now clear that the late-time increase in $c_s^2$ is crucial
for the strong time variation of the gravitational potential that leads to the large ISW signal.
Therefore, to avoid enhancement of the ISW effect,
$c_s^2$ must be sufficiently small {\em almost all the way from the past to the present}.
This point has already been emphasized in~\cite{Bertacca:2008uf} (see
also \cite{Piattella:2009kt}).
For sufficiently small but finite $c_s^2$,
weak gravitational lensing can be used to discriminate among
UDM models~\cite{Camera:2009uz} because small $c_s^2$ pushes the Jeans length away to small scales.
In order to give a reliable prediction about weak lensing,
it is important to address the nonlinear evolution of perturbations
in UDM models.
This issue would be very interesting and is left for further study.

Let us finally comment on the limiting case $\alpha \rightarrow \infty$. 
In this limit, the (scalar field model of) gCg behaves like Cuscuton proposed in
\cite{Afshordi:2006ad}. While the Cuscuton has the infinite sound
speed, it does not transport any information. It would be very
interesting to explore the consequence of this exceptional case \cite{Piattella:2009da}.

\acknowledgments
Y.U. and T.K. are grateful to O. Piattella, A. Starobinsky,
and H. Tashiro for their variable comments. The authors also would like to thank
K. Maeda for his continuous encouragement.
Y.U. and T.K. are supported
by the JSPS under Contact Nos. 19-720 and 19-4199. 


\end{document}